\documentclass[useAMS,usenatbib,usegraphicx]{mn2e}

\def\aj{AJ}%
\def\araa{ARA\&A}%
\def\apj{ApJ}%
\def\apjl{ApJ}%
\def\apjs{ApJS}%
\def\apss{Ap\&SS}%
\def\aap{A\&A}%
\def\aaps{A\&AS}%
\def\mnras{MNRAS}%
\def\pasp{PASP}%
\def\nat{Nature}%
\title[The awakening of BL Lacertae]
{The awakening of BL Lacertae:  
observations by {\em Fermi}, {\em Swift}, and the GASP-WEBT
\thanks{The radio-to-optical data collected by the GASP-WEBT collaboration are stored in the GASP-WEBT archive; for questions regarding their availability, please contact the WEBT President Massimo Villata ({\tt villata@oato.inaf.it}).}}
\author[C. M. Raiteri et al.] 
{C.~M.~Raiteri            ,   $^{ 1}$\thanks{E-mail:raiteri@oato.inaf.it}
M.~Villata               ,   $^{ 1}$
F.~D'Ammando             ,   $^{ 2,3}$
V.~M.~Larionov           ,   $^{ 4,5,6}$
\newauthor
M.~A.~Gurwell            ,   $^{ 7}$
D.~O.~Mirzaqulov         ,   $^{ 8}$
P.~S.~Smith              ,   $^{ 9}$
J.~A.~Acosta-Pulido      ,   $^{10,11}$
\newauthor
I.~Agudo                 ,   $^{12,13}$
M.~J.~Ar\'evalo          ,   $^{10,11}$
R.~Bachev                ,   $^{14}$
E.~Ben\'itez             ,   $^{15}$
A.~Berdyugin             ,   $^{16}$
\newauthor
D.~A.~Blinov             ,   $^{ 4,17}$
G.~A.~Borman             ,   $^{18}$
M.~B\"ottcher            ,   $^{19,20}$
V.~Bozhilov              ,   $^{21}$
\newauthor
M.~I.~Carnerero          ,   $^{ 1,10,11}$
D.~Carosati              ,   $^{22,23}$
C.~Casadio               ,   $^{12}$
W.~P.~Chen               ,   $^{24}$
\newauthor
V.~T.~Doroshenko         ,   $^{25}$
Yu.~S.~Efimov            ,   $^{18}$
N.~V.~Efimova            ,   $^{ 4,5}$
Sh.~A.~Ehgamberdiev      ,   $^{ 8}$
\newauthor
J.~L.~G\'omez            ,   $^{12}$
P.~A.~Gonz\'alez-Morales ,   $^{10}$
D.~Hiriart               ,   $^{15}$
S.~Ibryamov              ,   $^{14}$
\newauthor
Y.~Jadhav                ,   $^{20}$
S.~G.~Jorstad            ,   $^{13,4}$
M.~Joshi                 ,   $^{13}$
V.~Kadenius              ,   $^{16}$
S.~A.~Klimanov           ,   $^{ 5}$
\newauthor
M.~Kohli                 ,   $^{20}$
T.~S.~Konstantinova      ,   $^{ 4}$
E.~N.~Kopatskaya         ,   $^{ 4}$
E.~Koptelova             ,   $^{24,26}$
\newauthor
G.~Kimeridze             ,   $^{27}$
O.~M.~Kurtanidze         ,   $^{27,28}$
E.~G.~Larionova          ,   $^{ 4}$
L.~V.~Larionova          ,   $^{ 4}$
\newauthor
R.~Ligustri              ,   $^{29}$
E.~Lindfors              ,   $^{16,32}$
A.~P.~Marscher           ,   $^{13}$
B.~McBreen               ,   $^{30}$
I.~M.~McHardy            ,   $^{31}$
\newauthor
Y.~Metodieva             ,   $^{21}$
S.~N.~Molina             ,   $^{12}$
D.~A.~Morozova           ,   $^{ 4}$
S.~V.~Nazarov            ,   $^{18}$
\newauthor
M.~G.~Nikolashvili       ,   $^{27}$
K.~Nilsson               ,   $^{32}$
D.~N.~Okhmat             ,   $^{18}$
E.~Ovcharov              ,   $^{21}$
\newauthor
N.~Panwar                ,   $^{24}$
M.~Pasanen               ,   $^{16}$
S.~Peneva                ,   $^{14}$
J.~Phipps                ,   $^{20}$
N.~G.~Pulatova           ,   $^{18,33}$
\newauthor
R.~Reinthal              ,   $^{16}$
J.~A.~Ros                ,   $^{34}$
A.~C.~Sadun              ,   $^{35}$
R.~D.~Schwartz           ,   $^{36}$
E.~Semkov                ,   $^{14}$
\newauthor
S.~G.~Sergeev            ,   $^{18}$
L.~A.~Sigua              ,   $^{27}$
A.~Sillanp\"a\"a         ,   $^{16}$
N.~Smith                 ,   $^{37}$
K.~Stoyanov              ,   $^{14}$
\newauthor
A.~Strigachev            ,   $^{14}$
L.~O.~Takalo             ,   $^{16}$
B.~Taylor                ,   $^{13,38}$
C.~Thum                  ,   $^{39}$
I.~S.~Troitsky           ,   $^{ 4}$
\newauthor
A.~Valcheva              ,   $^{21}$
A.~E.~Wehrle             ,   $^{40}$
and H.~Wiesemeyer               $^{41,39}$\\
$^{ 1}$INAF, Osservatorio Astrofisico di Torino, Italy                                                    
$^{ 2}$Dip.\ di Fisica, Universit\`a degli Studi di Perugia, Perugia, Italy                                \\
$^{ 3}$INAF, Istituto di Radioastronomia, Bologna, Italy                                                   
$^{ 4}$Astron.\ Inst., St.-Petersburg State Univ., Russia                                                  \\
$^{ 5}$Pulkovo Observatory, St.-Petersburg, Russia                                                         
$^{ 6}$Isaac Newton Institute of Chile, St.-Petersburg Branch                                              \\
$^{ 7}$Harvard-Smithsonian Center for Astrophysics, Cambridge, MA, USA                                     \\
$^{ 8}$Maidanak Observatory of the Ulugh Beg Astronomical Institute, Uzbekistan                            \\
$^{ 9}$Steward Observatory, University of Arizona, Tucson, AZ, USA                                         
$^{10}$Instituto de Astrofisica de Canarias (IAC), La Laguna, Tenerife, Spain                              \\
$^{11}$Departamento de Astrofisica, Universidad de La Laguna, La Laguna, Tenerife, Spain                   \\
$^{12}$Instituto de Astrof\'{i}sica de Andaluc\'{i}a, CSIC, Granada, Spain                                 
$^{13}$Institute for Astrophysical Research, Boston University, MA, USA                                    \\
$^{14}$Institute of Astronomy, Bulgarian Academy of Sciences, Sofia, Bulgaria                              \\
$^{15}$Instituto de Astronom\'ia, Universidad Nacional Aut\'onoma de M\'exico, Ensenada, M\'exico          \\
$^{16}$Tuorla Observatory, Dept.\ of Physics and Astronomy, Univ.\ of Turku, Piikki\"o, Finland            \\
$^{17}$Physics  Department, University of Crete, Heraklion, Greece                                         
$^{18}$Crimean Astrophysical Observatory, Ukraine                                                          \\
$^{19}$Centre for Space Research, North-West University, Potchefstroom, South Africa                       \\
$^{20}$Department of Physics and Astronomy, Ohio Univ., OH, USA                                            
$^{21}$Dept.\ of Astronomy, Faculty of Physics, Sofia University, Bulgaria                                 \\
$^{22}$EPT Observatories, Tijarafe, La Palma, Spain,                                                       
$^{23}$INAF, TNG Fundaci\'on Galileo Galilei, La Palma, Spain                                              \\
$^{24}$Graduate Inst.\ of Astronomy, National Central Univ., Jhongli, Taiwan                               \\
$^{25}$South Station of the Moscow MV Lomonosov State University, Moscow, Russia, Crimea, Ukraine          \\
$^{26}$Department of Physics, National Taiwan University, Taipei, Taiwan                                   
$^{27}$Abastumani Observatory, Mt. Kanobili, Abastumani, Georgia                                           \\
$^{28}$Landessternwarte Heidelberg-K\"onigstuhl, Heidelberg, Germany                                       
$^{29}$Circolo Astrofili Talmassons, Italy                                                                 \\
$^{30}$UCD School of Physics, University College Dublin, Dublin, Ireland                                   \\
$^{31}$Dept.\ of Physics and Astronomy, Univ.\ of Southampton, Southampton, United Kingdom                 \\
$^{32}$Finnish Centre for Astronomy with ESO (FINCA), University of Turku, Piikki\"o, Finland              \\
$^{33}$National Astronomical Research Institute of Thailand                                                
$^{34}$Agrupaci\'o Astron\`omica de Sabadell, Spain                                                        \\
$^{35}$Department of Physics, Univ.\ of Colorado Denver, CO, USA                                           
$^{36}$Galaxy View Observatory, Sequim, Washington, USA                                                    \\
$^{37}$Cork Institute of Technology, Cork, Ireland                                                         
$^{38}$Lowell Observatory, Flagstaff, AZ, USA                                                              \\
$^{39}$Instituto de Radio Astronom\'{i}a Milim\'{e}trica, Granada, Spain                                   \\
$^{40}$Space Science Institute, Boulder, CO, USA                                                           
$^{41}$Max-Planck-Institut f\"ur Radioastronomie, Bonn, Germany                                            \\
 }
\begin{document}
\maketitle
\clearpage
\begin{abstract}
Since the launch of the {\em Fermi} satellite, BL Lacertae has been moderately active at $\gamma$-rays and optical frequencies until May 2011, when the source started a series of strong flares.
The exceptional optical sampling achieved by the GLAST-AGILE Support Program (GASP) of the Whole Earth Blazar Telescope (WEBT) in collaboration with the Steward Observatory allows us to perform a detailed comparison with the daily $\gamma$-ray observations by {\em Fermi}.
Discrete correlation analysis between
the optical and $\gamma$-ray emission reveals correlation with a time lag of $0 \pm 1$ d, which suggests cospatiality of the corresponding jet emitting regions. A better definition of the time lag is hindered by 
the daily gaps in the sampling of the extremely fast flux variations.
In general, optical flares present more structure and develop on longer time scales than corresponding $\gamma$-ray flares.  Observations at X-rays 
and at millimetre wavelengths reveal a common trend, which suggests that the region producing the mm and X-ray radiation is located 
downstream from the optical and $\gamma$-ray-emitting zone in the jet.
The mean optical degree of polarisation slightly decreases over the considered period and in general it is higher when the flux is lower. The optical electric vector polarisation angle (EVPA) shows a preferred orientation of about 15\degr, nearly aligned with the radio core EVPA and mean jet direction. Oscillations around it increase during the 2011--2012 outburst. 
We investigate the effects of a geometrical interpretation of the long-term flux variability on the polarisation.
A helical magnetic field model predicts an evolution of the mean polarisation that is in reasonable agreement with the observations. These can be fully explained by introducing slight variations in the compression factor in a transverse shock waves model.
\end{abstract}


\section{Introduction}
BL Lacertae is the prototype of a class of active galactic nuclei (AGNs) that together with flat-spectrum radio quasars (FSRQs) make up the collection of highly variable objects known as ``blazars".
The common features of all blazars is to show strong flux and spectral variability at all wavelengths and on a variety of time scales \citep{wag95}. 
They are also highly variable in optical and radio polarization \citep{smi96,all96}. 
The analysis of their radio map evolution reveals apparent superluminal motion of knots, which follow curved trajectories \citep{kel04}. It is believed that blazar emission comes from a relativistic plasma jet seen at a small angle to the line of sight, with consequent relativistic beaming of the radiation \citep{urr95}. The location of the emitting regions inside the jet and the structure of the jet itself are still a matter of debate. The observed low-energy radiation (from radio to UV or even X-rays in some sources) is due to synchrotron emission from relativistic electrons, which can also produce high-energy (X- and $\gamma$-ray) photons through an inverse-Compton mechanism \citep{kon81}. Cross-correlation analysis between flux variations in different bands can allow us to establish whether the emissions come from the same region in the jet \citep{huf92}, and in the case that they do not, give an indication of the relative distance of the emitting zones. 

To gain insight in the blazar properties, multifrequency campaigns are organised, involving many ground-based observatories as well as satellite observations \citep[e.g.][]{mar10,jor10,agu11a,agu11b}. The Whole Earth Blazar Telescope (WEBT)\footnote{{\tt http://www.oato.inaf.it/blazars/webt/}} was born in 1997 to study specific objects over a limited period of time \citep[e.g.][and references therein]{vil07,rai08a,lar08,boe09}.

Ten years later, the WEBT started the GLAST-AGILE Support Program (GASP), with the aim of performing low-energy monitoring of a selected sample of 28 blazars to compare with the high-energy observations of the $\gamma$-ray satellites Astrorivelatore Gamma ad Immagini LEggero \citep[AGILE;][]{tav09} and {\em Fermi}\footnote{Formerly GLAST} \citep{abd09_fermi}.
Results obtained by the GASP have been reported in, e.g., \citet{vil08,vil09b} and \citet{rai11,rai12}.

BL Lacertae is a bright blazar at low redshift \citep[$z=0.069$,][]{mil77}, hosted by a giant elliptical galaxy with 
$R=15.5$ \citep{sca00}.
It has already been the subject of several multiwavelength studies carried out by the WEBT/GASP \citep{vil02,vil04a,vil04b,bac06,pap07,rai09,vil09a,lar10,rai10}. 
In this new paper we analyse the $\gamma$-ray, X-ray, UV, optical, and millimetric behaviour of BL Lacertae from the start of the GASP observations of this object, in early 2008, through the period of strong activity in 2011--2012, until 2012 October 31, 
when the source came back to an optical and $\gamma$-ray ``quiescent" state.
In a forthcoming paper we will study the optical-to-radio historical flux and spectral behaviour of the source since the birth of the WEBT collaboration in 1997 (Raiteri et al, in preparation, Paper II).

\section{Optical photometry}
\label{optical}

\begin{figure*}
\centering
\includegraphics[width=15cm]{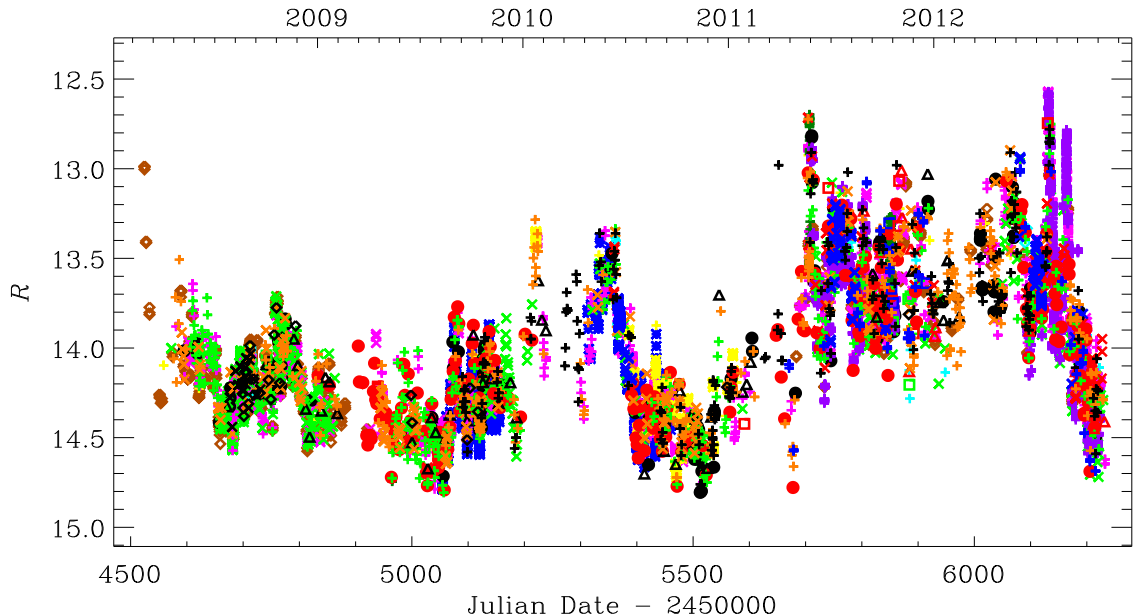}
\caption{Light curve of BL Lacertae in the $R$ band. The 10103 data points represent observed magnitudes, without correction for the host galaxy contribution and Galactic extinction. The various datasets are plotted with different colours and symbols to highlight the composite nature of the curve, requiring an accurate data assembling and checking process.}
\label{rmag}
\end{figure*}

The ground-based optical photometry presented in this paper was obtained by the GASP collaboration with the contribution of the Steward Observatory programme in support of the {\em Fermi} $\gamma$-ray telescope\footnote{\tt http://james.as.arizona.edu/$\sim$psmith/Fermi} \citep{smi09}.
Figure \ref{rmag} shows the $R$-band light curve of BL Lacertae from the start of the GASP observations of this source, on 2008 February 28 ($\rm JD=2454524.6$), up to 2012 October 31 ($\rm JD=2456232.3$).
The data points represent observed magnitudes, with no correction for the Galactic extinction and host-galaxy contribution (see below).
Calibration of the source magnitude was performed with respect to Stars B, C, and H by \citet{fio96}.

Data up to $\rm JD \sim 2454900$ have partially been presented in \citet{rai10}.
The new observations featured in this paper were provided by the following observatories: 
Abastumani,           
AstroCamp,                 
Belogradchik,                 
Calar Alto\footnote{Calar Alto data was acquired as part of the MAPCAT project: http://www.iaa.es/$\sim$iagudo/research/MAPCAT},                   
Crimean,  
Galaxy View,                   
Kitt Peak (MDM),   
Lowell (Perkins),     
Lulin,                   
Mt. Maidanak,      
New Mexico Skies,             
ROVOR,                         
Roque de los Muchachos (KVA and Liverpool),          
Rozhen,              
Sabadell,                   
San Pedro Martir,    
Skinakas,                     
St. Petersburg,  
Steward (Bok and Kuiper),              
Talmassons,                     
Teide (IAC80),                     
and Tijarafe.

\begin{figure*}
\centering
\includegraphics[width=12cm]{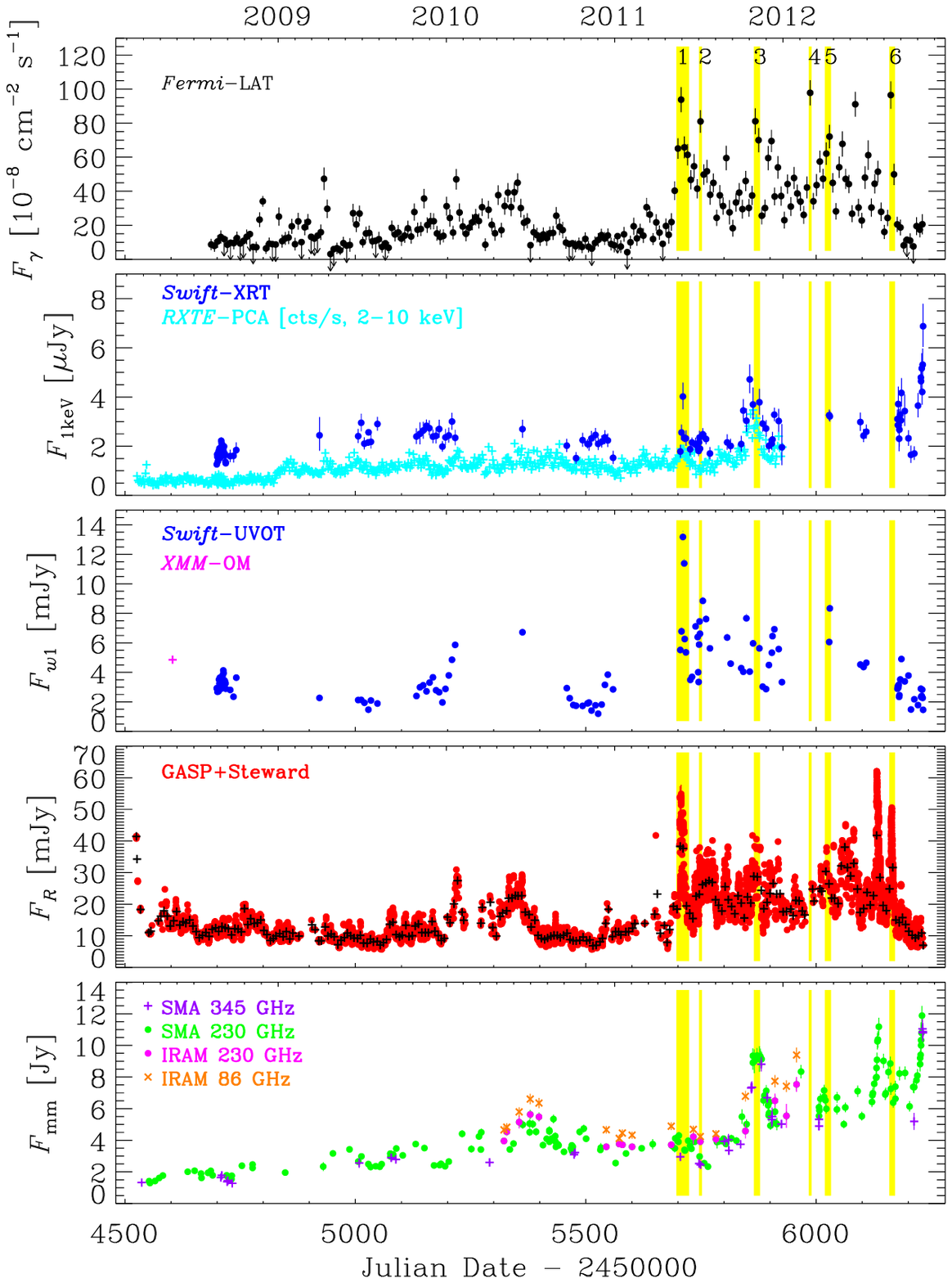}
\caption{From top to bottom: 
a) Integrated flux light curve of BL Lacertae in the 0.1--100 GeV energy range obtained during 2008 August 4 -- 2012 October 31 with 7-day time bins. Arrows refer to 2-$\sigma$ upper limits on the source flux. 
b) {\em Swift}-XRT unabsorbed flux densities at 1 keV (blue dots) and {\em RXTE}-PCA count rate in the 2--10 keV range (cyan plus signs). 
c) {\em Swift}-UVOT unabsorbed flux densities in the $w1$ band;  
one observation performed by {\em XMM}-Newton in 2008 is also shown \citep[from][]{rai10}. 
d) $R$-band flux densities obtained from the magnitudes shown in Fig.\ \ref{rmag} after correction for Galactic extinction and host galaxy contribution. The black crosses represent the result of a weekly binning. e) Millimetre light curve built with data at 230 GHz (green dots) and 345 GHz (violet plus signs) acquired at the SMA as well as data taken at 230 GHz (purple dots) and 86 GHz (orange crosses) with the 30-m IRAM telescope on Pico Veleta.
In all panels the yellow stripes indicate the periods considered for the $\gamma$-ray spectral analysis in Sect.\ \ref{fermi}.}
\label{multi}
\end{figure*}

The light curve shown in Fig.\ \ref{rmag} was obtained after a careful analysis, where the different datasets were assembled, checked and cleaned for offsets and outliers (a detailed description of the process will be given in Paper II), and includes 10103 data points.
Offsets caused by partial inclusion of the host galaxy were minimised by adopting the same prescriptions for the photometry: an aperture radius of 8\arcsec\ for the source and reference stars, and an annulus of 10\arcsec\ and 16\arcsec\ radii centred on them for the background. 
This choice of standard aperture includes 60\% of the total flux from the host galaxy, corresponding to a flux density of 2.54 mJy in the Cousins' $R$ band \citep{rai10}. 
The $R$-band flux densities of BL Lacertae, corrected for both a Galactic extinction of 0.88 mag and the above host-galaxy contribution, are shown in Fig.\ \ref{multi}.

Strong variability characterises the entire period on a large variety of time scales.
In particular, the outburst of 2011--2012 appears as a period of about 500 days where the source magnitude oscillated in the range $R = 12.57$--14.31. Very rapid flux changes can occur on time scales of much less than a day, as already noticed in previous works \citep[e.g.][]{rai09}. A detailed analysis of variability on the whole optical light curves built with GASP-WEBT observations will be performed in Paper II.

\section{Millimetre observations}

Millimetre observations were performed at the Submillimeter Array (SMA) and at the IRAM 30-m telescope\footnote{IRAM 30 m data were acquired as part of the POLAMI (Polarimetric AGN Monitoring with the IRAM-30 m-Telescope) and MAPI (Monitoring AGN with Polarimetry at the IRAM-30m- Telescope) programmes.}.

Data at 230 GHz (1.3 mm) and 345 GHz (870 micron) were obtained at the SMA near the summit of Mauna Kea (Hawaii). BL Lacertae is included in an ongoing monitoring program at the SMA to determine the fluxes of compact extragalactic radio sources that can be used as calibrators at mm/submm wavelengths \citep{gur07}. Available potential calibrators are observed for 3 to 5 minutes, and the measured source signal strength calibrated against known standards, typically solar system objects (Titan, Uranus, Neptune, or Callisto). Despite the short integration time, the flux calibration error is dominated by systematic effects such as pointing or phase instability, for sources greater than about 250 mJy, such as BL Lacertae. Data from regular science tracks are also reduced to obtain flux measurements from time to time, and these data often are taken over several hours. Data from this monitoring program are updated regularly and are available at the SMA website\footnote{\tt http://sma1.sma.hawaii.edu/callist/callist.html}. BL Lac was also observed as part of two dedicated programs to monitor its flux density (PI: A. Wehrle), and data from those programs through 2012 October 31 are included here. 

The IRAM 30 m telescope (in Granada, Spain) observed simultaneously at 86.24 GHz (3.5 mm), and 228.93 GHz (1.3 mm) by making use of the EMIR090, and EMIR230 pairs of orthogonally linearly polarized heterodyne receivers connected to the XPOL photo-polarimeter \citep{thu08}.
For our observations, a bandwidth of 640 MHz was used for each of the EMIR090 receivers, whereas 260 MHz were used for the 228.93 GHz measurements with EMIR230.
Every IRAM 30 m measurement was preceded by a cross-scan pointing of the telescope and a 3.5 mm and a 1.3 mm calibration.
Such measurements consisted on series of wobbler switching on-offs with total integration times of 4 min to 8 min, depending on the total flux density of the source and atmospheric conditions.
Measurements of Mars and/or Uranus were obtained, at least once, essentially for every observing epoch in order to estimate and subtract residual instrumental polarization, and to calibrate the absolute total flux density scale. 
Data reduction was performed following the procedures described in \citet{agu06,agu10}.
The resulting data were averaged for those observing epochs on which more than one measurement was obtained.

The mm light curve at 230 GHz built with IRAM and SMA data, as well as the light curves at 86 GHz from IRAM and at 345 GHz from the SMA are plotted in the bottom panel of Fig.\ \ref{multi}.

\section{{\em Swift}-UVOT}

The {\em Swift} satellite carries a 30-cm Ultraviolet/Optical telescope (UVOT; \citealt{rom05}) that can acquire data in 6 filters: $v$, $b$, and $u$ in the optical band; $uvw1$, $uvm2$, and $uvw2$ in the ultraviolet.  
We reduced the BL Lac observations with the {\texttt {HEASoft}} package version 6.12 and the 20120606 release of the Swift/UVOTA calibration database {\texttt {CALDB}} at NASA's High Energy Astrophysics Science Archive Research Center (HEASARC)\footnote{\texttt {http://heasarc.nasa.gov/}}.
Multiple observations in the same filter at the same epoch were first summed with the task {\texttt {uvotimsum}} and then processed with {\texttt {uvotsource}}. Source counts were extracted from a circular region centred on the source with 5\arcsec\ radius. Background counts were derived from an annular region centred on the source with 10\arcsec\ and 16\arcsec\ radii.

\begin{figure*}
\centering
\includegraphics[width=12cm]{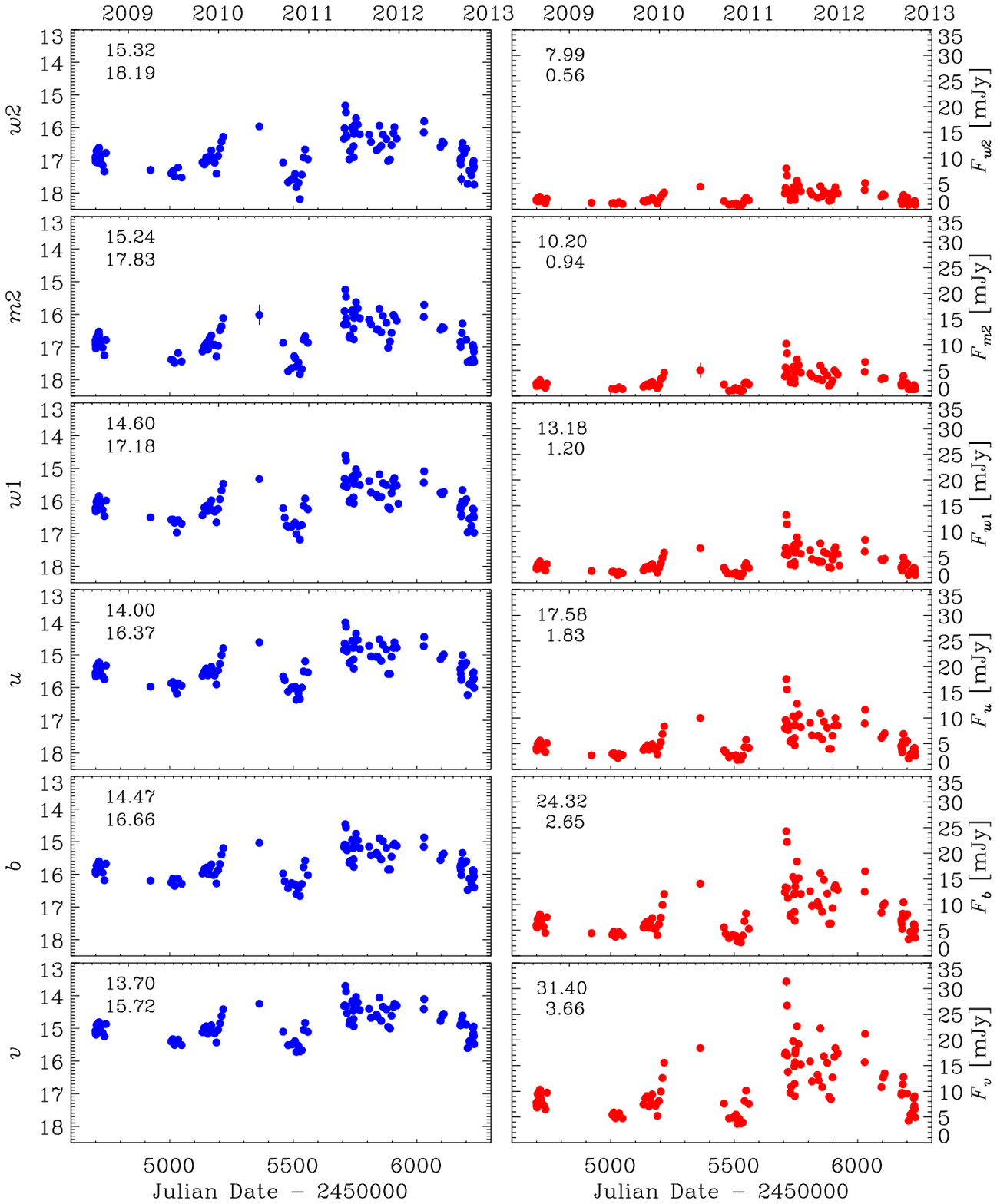}
\caption{Light curves of BL Lacertae at optical and UV frequencies obtained from the observations of the UVOT instrument onboard the {\em Swift} satellite. Left: Observed magnitudes. Right: Flux densities in mJy, after correction for the Galactic extinction and subtraction of the host-galaxy contribution. In all plots the numbers in the upper left indicate the maximum and minimum brightness levels.}
\label{uvot}
\end{figure*}

The UVOT light curves in the period considered here are shown in Fig.\ \ref{uvot}, both as observed magnitudes (left) and as intrinsic flux densities (right).
The latter have been obtained by correcting for the Galactic extinction and by subtracting the host galaxy contribution, as explained below.
We calculated a Galactic extinction of 1.09, 1.44, 1.73, 2.52, 3.05, and 2.91 mag in the $v$, $b$, $u$, $uvw1$, $uvm2$, and $uvw2$ bands, respectively, by convolving the new effective areas of the UVOT filters by \citet{bre11} with the mean Galactic extinction laws by \citet{car89}.  
Following \citet{rai10} we assumed a flux density of 2.89, 1.30, and 0.36 mJy for the host galaxy in the $v$, $b$, and $u$ bands. In the UV, we considered the 13 Gyr elliptical galaxy spectral template by \citet{sil98}, and estimated a host galaxy flux density of 0.026, 0.020, and 0.017 mJy in the $uvw1$, $uvm2$, and $uvw2$ bands.
The host galaxy contribution contaminating the BL Lac UVOT photometry is about 50\% of the whole galaxy flux, so in the UV it is negligible when compared to the source flux, even in faint states.

As can be seen from Fig.\ \ref{uvot}, the variability amplitude increases with frequency: the difference between the maximum and minimum magnitude is 
2.02, 2.20, 2.37, 2.59, 2.59, and 2.87 mag from the $v$ to the $uvw2$ band, confirming the behaviour already observed in this source \citep[e.g.][]{rai10} and, in general, in BL Lac objects.
Notice that this trend does not depend on the presence of the host galaxy, because it remains also when the host galaxy contribution is removed. Indeed, the ratio between the maximum and minimum intrinsic flux density is 8.6, 9.2, 9.6, 11.0, 10.9, and 14.3 going from the $v$ to the $uvw2$ band.
Moreover, while in FSRQ the presence of thermal radiation from the accretion disc can imply smaller flux variability toward the UV, here its likely contribution \citep{rai09,rai10,cap10} is probably not strong enough to contrast the typical behaviour of the synchrotron emission.

The unabsorbed $uvw1$ flux densities are also plotted in Fig.\ \ref{multi} for a comparison with other bands. 

\section{{\em Swift}-XRT}

We processed the X-ray Telescope (XRT; \citealt{bur05}) data with the {\texttt {HEASoft}} package version 6.12 and the CALDB XRT calibration files updated 20120209.
The task {\texttt {xrtpipeline}} was launched with standard screening criteria.
Only observations performed in pointing mode and with more than 50 counts were selected for further analysis. 
In the time period considered in this paper, we were left with 117 observations in photon counting (PC) mode. 

\begin{figure}
\centering
\resizebox{\hsize}{!}{\includegraphics{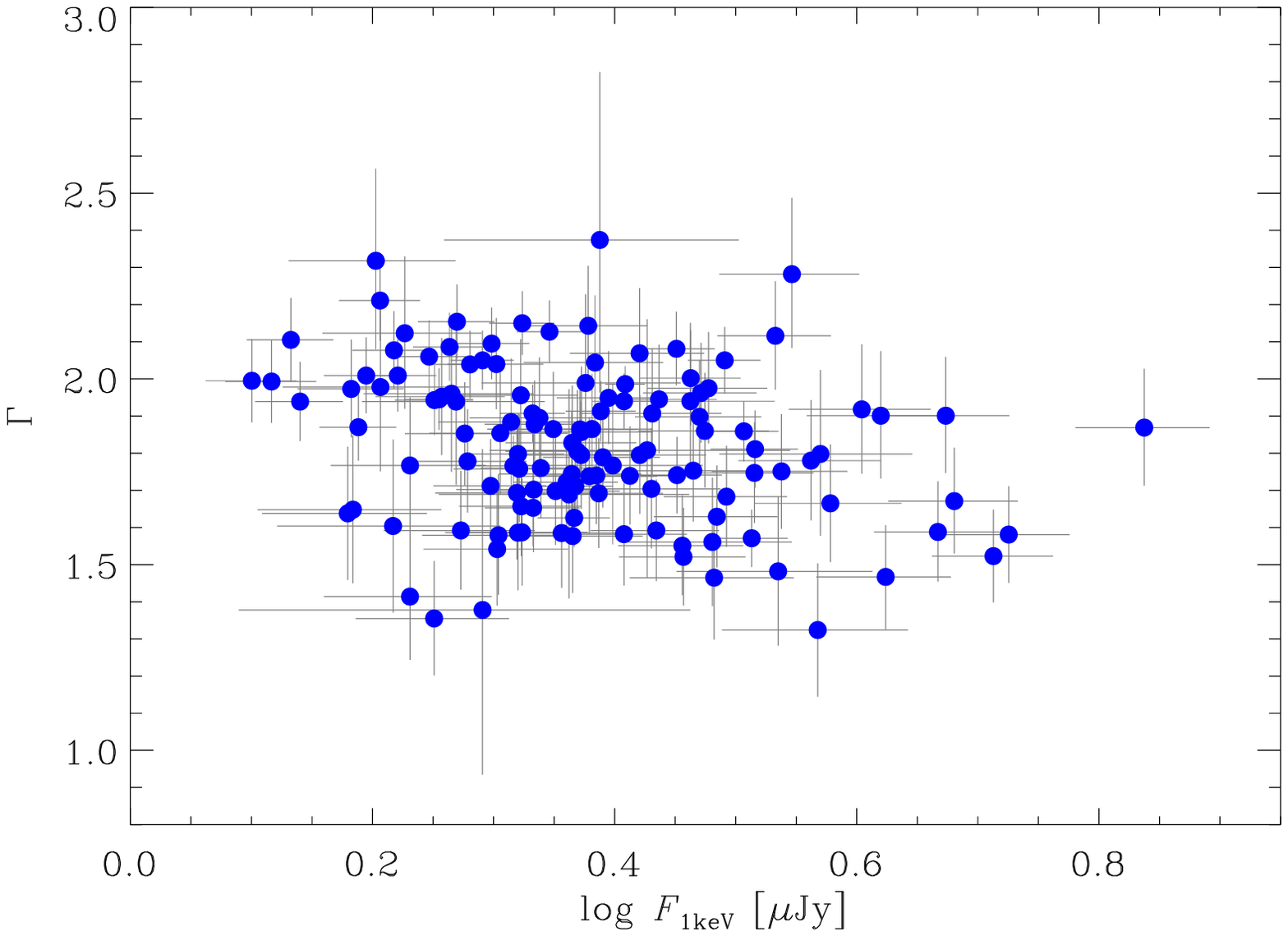}}
\caption{The X-ray photon index $\Gamma$ as a function of the unabsorbed flux density at 1 keV when a power-law model with Galactic absorption fixed to $N_{\rm H} = 3.4 \times 10^{21} \, \rm cm^{-2}$ is applied to the XRT spectra.}
\label{gamma}
\end{figure}

All PC observations with mean rate greater than 0.5 cts/s were checked for pile-up with the task {\texttt {XIMAGE}}\footnote{\texttt {http://www.swift.ac.uk/analysis/xrt/pileup.php}}. The wings of the source point spread function (PSF) were modeled with the expected PSF of XRT, i.e.\ a King function of the type: ${\rm PSF} (r) = [ 1 + (r/5.8)^2 ]^{-1.55}$  \citep{mor05}. The fit was then extrapolated to the inner region and compared to the data points. The radius below which the model overproduces the data defines the region where pile-up is a problem.

Source counts were extracted with the {\texttt {xselect}} task from a circular region of 30 pixel (71\arcsec) radius centred on the source, and background counts from a surrounding annulus of 50 and 70 pixel radii, respectively.
For piled-up observations, we excluded from the source extraction region the inner circle of 3 pixel radius ($\sim 7 \arcsec$).

The loss of counts caused by the inner hole in the source counts extraction region is corrected by the ancillary response file, which also takes account of vignetting and bad pixels.
This file is obtained through the {\texttt {xrtmkarf}} task with PSF correction set to yes and using the exposure map created by {\texttt {xrtpipeline}}. We adopted version 011 of the response matrix available in {\texttt {CALDB}}.

The source spectra were grouped with the task {\texttt {grppha}} and then analysed in the 0.3--10 keV energy range with the {\texttt {Xspec}} task, using both the Cash and $\chi^2$ statistics. In the latter case, the spectra were previously binned to have a minimum of 20 counts in each bin. 
Spectra were fitted with an absorbed power law. Following \citet{rai09,rai10}, we adopted a Galactic hydrogen column density (including the contribution by a molecular cloud toward BL Lac, see \citealt{lis98})  of $N_{\rm H} = 3.4 \times 10^{21} \, \rm cm^{-2}$ and set abundances for photoelectric absorption according to \citet{wil00}.

Figure \ref{gamma} displays the photon index $\Gamma$ as a function of the flux density at 1 keV. The $\Gamma$ values are scattered between 1.32 (hard spectrum) and 2.37 (soft spectrum) without correlation with the flux.
The XRT data (unabsorbed flux densities at 1 keV) are plotted in Fig.\ \ref{multi}, where they are compared to the Rossi X-ray Timing Explorer {\em RXTE} Proportional Counter Array (PCA) light curve (cts/s in the 2--10 keV energy range) publicly available through the ISDC-HEAVENS interface\footnote{http://www.isdc.unige.ch/heavens/}.

\section{{\em Fermi}-LAT}
\label{fermi}

The {\em Fermi}-LAT is a pair-conversion telescope operating from 20 MeV to
$>$ 300 GeV. It has a large peak effective area ($\sim$ 8000 cm$^{2}$ for 1
GeV photons), an energy resolution of typically $\sim$10\%, and a field of
view of about 2.4 sr with an angular resolution (68\% containment angle)
better than 1$^{\circ}$ for energies above 1 GeV. Further details about the {\em Fermi}-LAT are given in \citet{atw09}.

The LAT data reported in this paper were collected from 2008 August 4  ($\rm JD=2454683$) to 2012 October 31 ($\rm JD=2456232$). During this time the {\em Fermi} spacecraft operated almost entirely in survey mode. The analysis was performed
with the \texttt{ScienceTools} software package version v9r27p1. The LAT data
were extracted within a $10^{\circ}$ Region of Interest (RoI) centred at the
radio location of BL Lacertae. Only events belonging to the ``Source'' class were used. In addition, a cut on the zenith angle ($< 100^{\circ}$) was applied to reduce contamination from the Earth limb $\gamma$-rays, which are produced by cosmic rays interacting with the upper atmosphere. 
The spectral analysis was performed with the instrument response functions 
\texttt{P7SOURCE\_V6} using an unbinned maximum likelihood method implemented
in the Science tool \texttt{gtlike}. A Galactic diffuse emission model and isotropic component, which is the sum of an extragalactic and instrumental background, were used to model the background\footnote{\texttt{http://fermi.gsfc.nasa.gov/ssc/data/access/lat/Background\\Models.html}}. The normalizations of both components in the background model were allowed to vary freely during the spectral fitting.

We evaluated the significance of the $\gamma$-ray signal from the sources by
means of the Test Statistics TS = 2$\Delta$log(likelihood) between models with
and without the source \citep{mat96}. For the spectral modelling of BL Lacertae we adopted a log-parabola, ${\rm d}N/{\rm d}E \propto (E/E_{0})^{-\alpha-\beta \, \log(E/E_0)}$
\citep[]{lan86, mas04}, as done in the 2FGL catalogue \citep{nol12}. The source model used in \texttt{gtlike} includes all the point sources from the 2FGL catalogue that fall within $20^{\circ}$ from our target. 
The spectra of these sources were parametrized by power-law functions, ${\rm d}N/{\rm d}E
\propto$ $(E/E_{0})^{-\Gamma}$, except for 2FGL J2111.3$+$4605, 2FGL
J2117.5$+$3730, 2FGL J2139.8$+$4714, 2FGL J2215.7$+$5135, and 2FGL
J2236.4$+$2828, for which we used a log-parabola as in the 2FGL catalogue. A
first maximum likelihood was performed to remove from the model the sources
having TS $<$ 25 and/or the predicted number of counts based on the fitted
model $N_{\rm pred} < 10$. 
A second maximum likelihood was performed on the updated source model. 
In the fitting procedure both the normalization factors and the photon indices of the sources within 10$^{\circ}$ from BL Lac were left as free parameters. For the sources located between 10$^{\circ}$ and 20$^{\circ}$ we kept the normalization and the photon index fixed to the values of the 2FGL catalogue.

\begin{figure*}
\centering
\includegraphics[width=15cm]{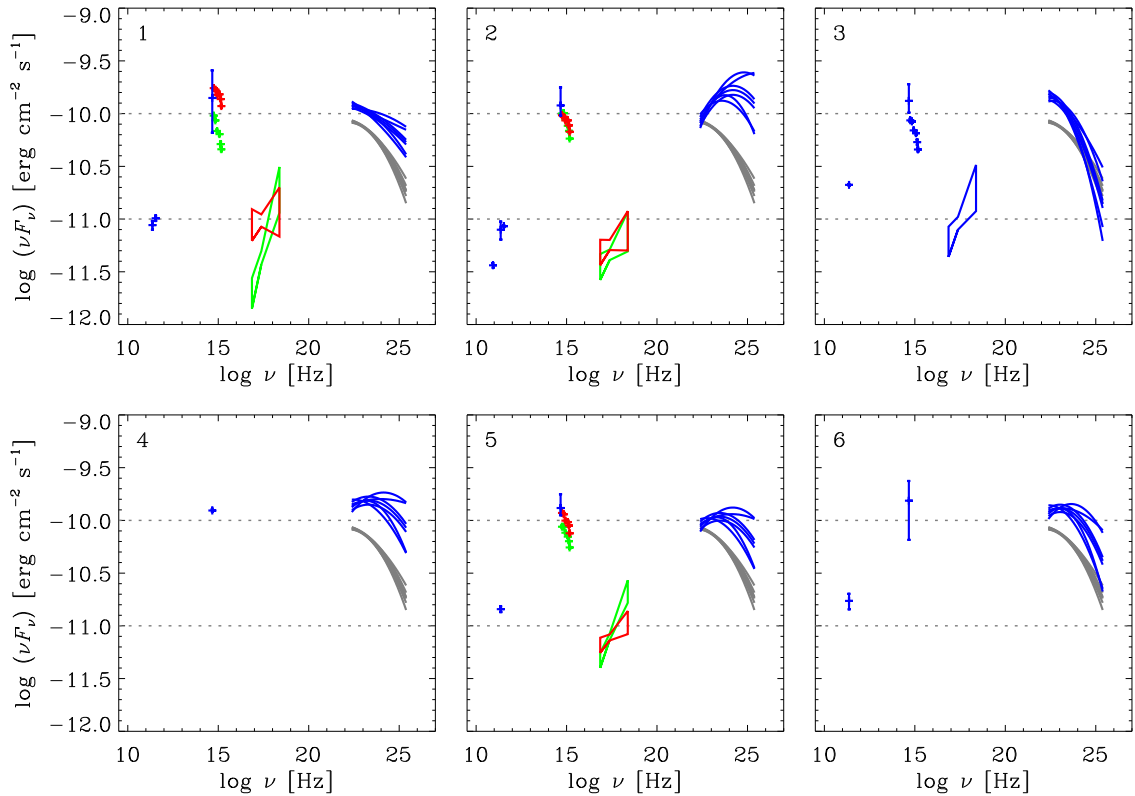}
\caption{Spectral energy distribution of BL Lacertae from the millimetre band to the $\gamma$-rays for the periods listed in Table \ref{gamspe}.
In each panel the period number is indicated in the upper left, and the $\gamma$-ray spectrum of period 0, including the whole outburst, is plotted in grey as a reference.
The spectral model used in $\gamma$-rays is a log-parabola ${\rm d}N/{\rm d}E \propto$ $(E/E_{0})^{-\alpha-\beta \, \log(E/E_0)}$, with the reference energy $E_0$ fixed to 388.5 MeV as in the 2FGL catalogue. 
The dispersion in the $\gamma$ spectrum is the consequence of plotting fits obtained with upper and lower limits of normalisation and of the parameters $\alpha$ and $\beta$ to display the uncertainties on both flux and spectral shape.
When contemporaneous {\em Swift} data are available, we show the corresponding X-ray power-law spectrum and optical--UV data or, in case there are multiple observations, the faintest (green) and brightest (red) states. In the $R$ band, there are usually many data within each considered period, hence we plot the whole range of flux values.
Millimetre data are also available for all epochs but one.
}
\label{LAT_spectra}
\end{figure*}

\begin{table*}
\begin{minipage}{126mm}
\caption{Results of the spectral analysis of the {\em Fermi}-LAT data in the 0.1--100 GeV energy range.
The fitting model is a log-parabola ${\rm d}N/{\rm d}E \propto$ $(E/E_{0})^{-\alpha-\beta \, \log(E/E_0)}$, with the reference energy $E_0$ fixed to 388.5 MeV as in the 2FGL catalogue.}
\label{gamspe}
\begin{tabular}{l l c c r c}
\hline
Period & Date & $\alpha$ & $\beta$ & TS & $F_{0.1-100 \, \rm GeV}$ \\
       &      &          &         &    & [$10^{-8} \rm \, ph \, cm^{-2} \, s^{-1}$]\\
\hline
0 & 2011 May 1 -- 2012 Aug 31 & 2.11$\pm$0.02 & 0.06$\pm$0.02 & 11354 & 47.0$\pm$0.9 \\
1 & 2011 May 15 -- Jun 11     & 2.08$\pm$0.06 & 0.02$\pm$0.02 & 1512  & 69.4$\pm$4.2 \\
2 & 2011 July 3--9            & 1.78$\pm$0.11 & 0.09$\pm$0.05 &  501  & 65.1$\pm$7.2 \\
3 & 2011 Oct 30 -- Nov 12     & 2.16$\pm$0.09 & 0.10$\pm$0.06 &  603  & 78.4$\pm$7.0 \\
4 & 2012 Feb 26 -- Mar 3      & 1.96$\pm$0.10 & 0.06$\pm$0.04 &  556  & 88.3$\pm$8.7 \\
5 & 2012 April 1--14          & 1.93$\pm$0.10 & 0.07$\pm$0.04 &  620  & 58.9$\pm$6.0 \\
6 & 2012 August 19--31        & 1.99$\pm$0.11 & 0.09$\pm$0.05 &  628  & 73.7$\pm$6.3 \\
\hline
\end{tabular}
\medskip
\end{minipage}
\end{table*}

Integrating over the entire period 2008 August 4 -- 2012 October 31 the fit
yielded TS = 13913 in the 0.1--100 GeV energy range, with an integrated
average flux of (25.8 $\pm$ 0.5) $\times$10$^{-8}$ photons cm$^{-2}$ s$^{-1}$,
a spectral slope $\alpha$ = 2.13 $\pm$ 0.02 at the reference energy $E_0$ =
388.5 MeV, and a curvature parameter around the peak $\beta$ = 0.07 $\pm$
0.01. 
The results of the spectral analysis for selected periods during the 2011--2012 outburst
are shown in Table \ref{gamspe} and Fig.\ \ref{LAT_spectra}. 
Period 0 includes all the outburst phase, from 2011 May 1 to 2012 August 31; the average apparent isotropic $\gamma$-ray luminosity in this period is $2.8 \times 10^{45} \rm \, erg \, s^{-1}$.
Periods from 1 to 6 were chosen by considering weekly bins with $\rm TS > 500$ or by summing subsequent 
bins with $\rm TS > 300$ and flux greater than $60 \times 10^{-8} \rm \, ph \, cm^{-2} \, s^{-1}$.

Figure~\ref{multi} shows the $\gamma$-ray light curve for the entire period
using a log-parabola model and 1-week time bins. For each time bin the
spectral parameters of BL Lacertae and all sources within 10$^{\circ}$ from it were frozen to the values resulting
from the likelihood analysis over the entire period. If TS $<$ 5, 
the values of the flux were replaced by the 2-$\sigma$ upper limits. The systematic
uncertainty on the flux is energy dependent: it amounts to $10\%$ at 100 MeV, decreasing to
$5\%$ at 560 MeV, and increasing to $10\%$ above 10 GeV
\citep{ack12}. By means of the \texttt{gtsrcprob} tool we estimated that
the highest-energy photon emitted by BL Lac was observed on 2012 March 9 at distance of
0.015$^{\circ}$ from the source with an energy of 74.3 GeV.

\begin{figure}
\centering
\resizebox{\hsize}{!}{\includegraphics{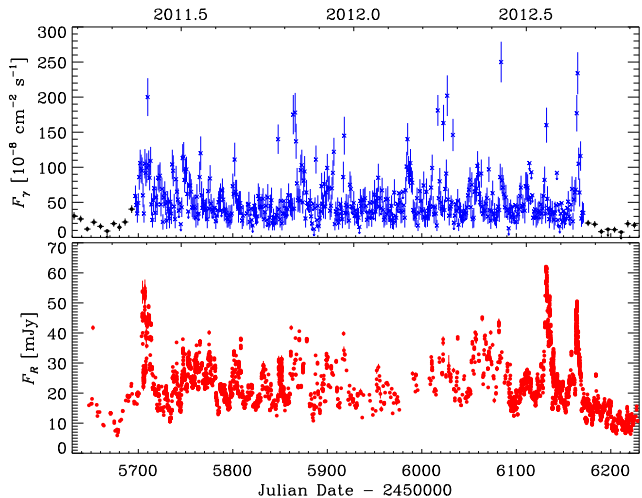}}
\caption{Top: Daily (blue crosses) and weekly (black dots) binned $\gamma$-ray light curve of BL Lacertae during the 2011--2012 outburst.
Bottom: Optical light curve in the same period.}
\label{multi_go_last}
\end{figure}

A second light curve focused on the period 2011 May 1 -- 2012 August 31 was
built with 1-day time bins. We used 12-hr and 6-hr time bins for the periods
with higher statistics. These daily and sub-daily light curves are shown in Figs.\ \ref{multi_go_last}--\ref{multi_go_zoc}, where they are compared with the optical flux. 

\section{Spectral energy distribution}
\label{sed}

The spectral energy distribution (SED) of BL Lacertae from the millimetric band to the $\gamma$-rays is shown in Fig.\ \ref{LAT_spectra} for the periods listed in Table \ref{gamspe} and highlighted by yellow stripes in Fig.\ \ref{multi}.
The {\em Fermi}-LAT spectrum corresponding to period 0, including the whole 2011--2012 outburst, is shown in all panels for comparison. The dispersion of the $\gamma$ spectra is due to the fact that besides the best fit we also show the fits obtained with the lower and upper limits on the flux normalisation, and on the parameters $\alpha$ and $\beta$. This illustrates the uncertainties involved. 

Simultaneous {\em Swift} data are available for periods 1, 2, 3, and 5. 
In particular, only one observation was performed in period 3, while two were done in periods 2 and 5 (see Fig.\ \ref{multi}). During period 1 there were six observations, so we show the XRT and UVOT spectra corresponding to the minimum and maximum flux levels.
We notice that the UV spectra do not show any hint for the presence of a UV bump, in contrast to what was found by \citet{rai10} during a low state. In that paper, we noticed that the UVOT calibration by \citet{poo08} was not suitable for a very red object like BL Lacertae, and performed a new calibration for this source. The result was an upturn of the spectrum in the UV, in agreement with data from the OM instrument onboard {\em XMM-Newton}.
The new calibration by \citet{bre11} implemented in the UVOT reduction software is not appropriate for very red objects too. Hence, in principle, we should proceed with a new re-calibration. However, the high state of BL Lacertae in 2011--2012 makes a search for a possible bump signature likely hopeless, so we neglected this point. 

Many $R$-band data were acquired in each considered period (apart from period 4, where just one datum is available because of the proximity to solar conjunction), showing large flux variation. In Fig.\ \ref{LAT_spectra} we plot the whole range of optical flux values.
Finally, millimetre observations were performed in all periods but one. 

The $\gamma$-ray spectrum of BL Lacertae shows a remarkable variability, suggesting that the inverse-Compton peak shifts from the MeV (periods 1 and 3) to the GeV (periods 2, 4, 5, and 6) range, but most of the time it is in the MeV domain (period 0). 
In period 1 the two UV and X-ray spectra are acquired on $\rm JD \approx 2455705.2$ (low state) and $\rm JD \approx 2455710.8$ (high state); they display a noticeable variability in both flux and spectral shape in only 5.6 d.
Moreover, we notice an optical--UV spectral steepening in the fainter state, corresponding to an X-ray spectral hardening.

\section{Comparison between $\gamma$-ray and optical flux variations.}

The results of the first 18 months of {\em Fermi} observations of BL Lacertae were presented by \citet{abd11_2200}. The source was in a low state, and no correlation between the $\gamma$-ray and optical fluxes was found.

If we compare the weekly-binned $\gamma$ and optical
light curves in Fig. \ref{multi}, we see that the ratio between the maximum and minimum flux level in $\gamma$-rays is about 15, while in the optical it is about 4, i.e.\ the $\gamma$-ray flux variability goes roughly as the square of the optical one. 
This is what is predicted by the synchrotron-self-Compton (SSC) theory for the origin of the $\gamma$-ray photons, according to which these are produced by inverse-Compton scattering of synchrotron photons created in the jet off their parent relativistic electrons \citep[e.g.][]{kon81,mar92}.

\begin{figure}
\centering
\resizebox{\hsize}{!}{\includegraphics{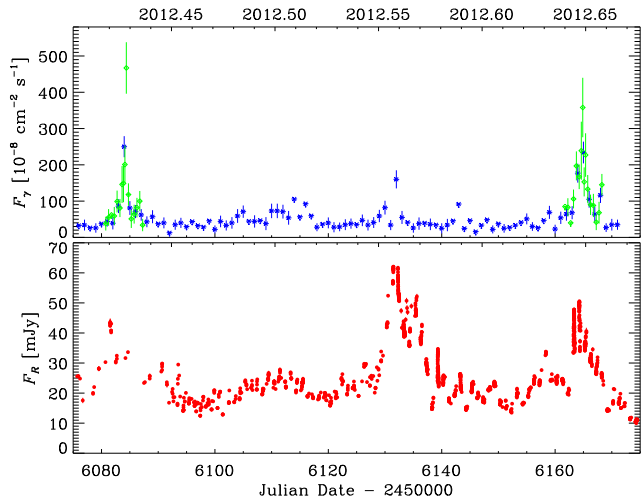}}
\caption{A zoom on the $\gamma$-ray (top) and optical (bottom) light curves at the culmination of the 2012 outburst, including the two strongest $\gamma$-ray flares. Sub-daily binned $\gamma$-ray fluxes (green diamonds) are superposed to the daily-binned ones (blue crosses).}
\label{multi_go_zo3}
\end{figure}

\begin{figure}
\centering
\resizebox{\hsize}{!}{\includegraphics{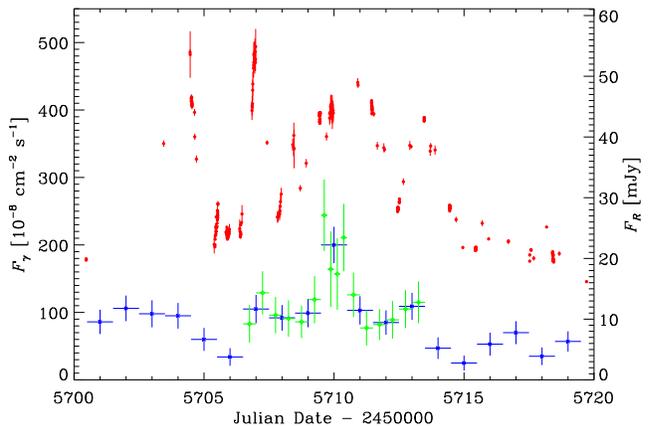}}
\caption{A comparison between the $R$-band flux densities (red dots) and $\gamma$-ray daily (blue crosses) and sub-daily (green diamonds) fluxes in 2011 May 18 -- June 7. }
\label{multi_go_zoa}
\end{figure}

\begin{figure}
\centering
\resizebox{\hsize}{!}{\includegraphics{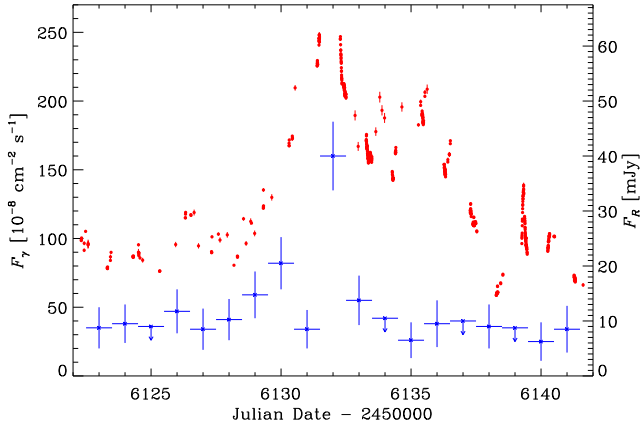}}
\caption{A comparison between the $R$-band flux densities (red dots) and $\gamma$-ray daily (blue crosses) fluxes in 2012 July 13 -- August 2.}
\label{multi_go_zob}
\end{figure}

\begin{figure}
\centering
\resizebox{\hsize}{!}{\includegraphics{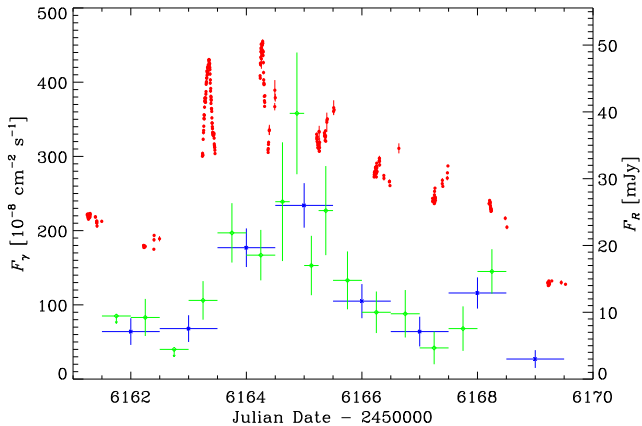}}
\caption{A comparison between the $R$-band flux densities (red dots) and $\gamma$-ray daily (blue crosses) and sub-daily (green diamonds) fluxes in 2012 August 21--30.}
\label{multi_go_zoc}
\end{figure}

The long-term $\gamma$-ray and optical light curves displayed in Fig.\ \ref{multi} confirm that in general the fluxes at these two frequencies are correlated. 
Most noticeably, the source clearly brightened in both $\gamma$ and optical bands after about 2011.3.
However, this correlation is not straightforward, as can be inferred from the more detailed Figs.\ \ref{multi_go_last}--\ref{multi_go_zoc}.
In particular, Fig.\ \ref{multi_go_zo3} shows the culmination of the 2012 outburst, with the highest $\gamma$-ray peaks, while Figs.\ \ref{multi_go_zoa}--\ref{multi_go_zoc} zoom into the periods of the major optical flares. 
We notice that the strongest observed $\gamma$-ray flare at $\rm JD=2456084$ does not correspond to the strongest observed optical flare, which peaked at $\rm JD=2456131$--$32$.
As for possible delays of flux variations in one band with respect to those in the other band, the situation around $\rm JD=2455710$ (Fig.\ \ref{multi_go_zoa}) appears confused, with many optical peaks either precedeing or following those in $\gamma$-rays.
An optical peak precedes the major $\gamma$-ray flare at $\rm JD=2456084$ by about three days (Fig.\ \ref{multi_go_zo3}), but the optical light curve is not sampled enough in that period to rule out that we missed a second optical flare closer in time to the $\gamma$ flare. 
The $\gamma$-ray and optical events at $\rm JD=2456131$--$32$ (Fig.\ \ref{multi_go_zob}) seem to be strictly simultaneous, assuming that the actual optical peak was missed and that we are only seeing the sharp wings of the optical flare. As for the flare at $\rm JD=2456163$--$64$ (Fig.\ \ref{multi_go_zoc}), the observed optical peaks appear to lead the $\gamma$-ray peak by at least 12 hours, but we could have missed an optical peak simultaneous to the $\gamma$-ray maximum.

\begin{figure}
\centering
\resizebox{\hsize}{!}{\includegraphics{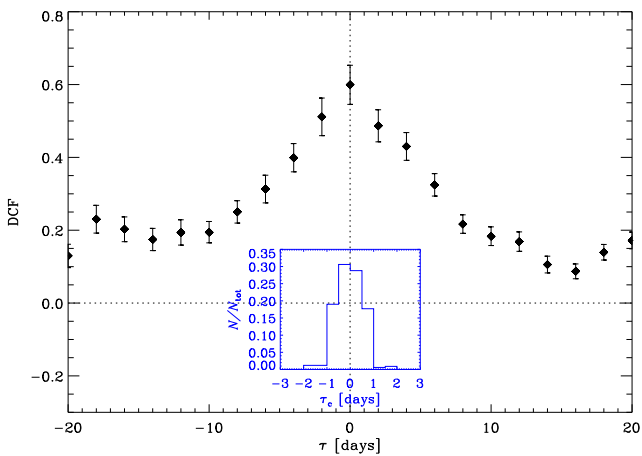}}
\caption{Discrete correlation function between the $\gamma$-ray fluxes and the $R$-band flux densities. The inset shows the result of cross-correlating 1000 Monte Carlo realisations of the two datasets according to the ``flux redistribution/random subset selection" technique.}
\label{dcf_go}
\end{figure}

We analysed the $\gamma$-optical cross-correlation with the discrete correlation function \citep[DCF;][]{ede88,huf92}, which was specifically designed for unevenly sampled datasets.
Figure \ref{dcf_go} shows the DCF obtained by cross-correlating a composite $\gamma$-ray light curve with the $R$-band flux densities. The composite $\gamma$-ray light curve includes weekly-binned data\footnote{Notice that the individual fluxes are associated with the central time of their bin.} before $\rm JD=2455697$ and daily or sub-daily-binned data afterwards.
Flux upper limits are substituted by data with half of the upper limit value and equal error. The DCF shows a well-defined peak at a time lag $\tau = 0 \, \rm d$, whose value is 0.60, indicating a fair correlation. The fact that the DCF peak is not higher may depend on the different ways the correlation reveals itself, as we saw above, as well as the different relative amplitude and duration of $\gamma$ and optical flares (see e.g.\ Fig.\ \ref{multi_go_zo3}).
The distribution of DCF values is roughly symmetric, which implies that the centroid $\tau_{\rm c}=(\sum_i \tau_i {\rm DCF}_i)/(\sum_i {\rm DCF}_i)$, where $i$ are all points with ${\rm DCF}_i$ close to the peak value, does not differ much from the time lag of the peak. To test the uncertainty of this result, we calculated the DCF for 1000 Monte Carlo realisations of the two datasets according to the ``flux redistribution/random subset selection" technique \citep{pet98,rai03}.
The inset of Fig.\ \ref{dcf_go} shows the fraction of simulations that resulted in a certain $\tau_{\rm c}$ bin. 
In this case, 96\% of simulations gave a time lag between $-1$ and +1 d. Although this is more than $1 \sigma$ uncertainty, it is not possible to reach a better resolution.  
In conclusion, the cross-correlation analysis seems to indicate that a correlation exists, even if it does not always show itself in the same way, and that the $\gamma$-ray flux variations can either follow (negative time lags) or precede (positive time lags) the optical fluctuations by 0--1 d in the observer's frame.

\section{Comparison between optical/$\gamma$ and mm/X-ray flux variations}

Figure \ref{multi} shows that the mm flux density is steadily increasing during the period considered in this paper, a feature that is not present in the optical and $\gamma$-ray light curves. 
A series of flares starts in late 2011, i.e.\ about 5 months after the beginning of the optical/$\gamma$-ray outburst.
The X-ray light curve is not well sampled in 2012, nevertheless a slow growth of the flux base level over the whole period can be recognised.
Some hints of flaring in the X-rays seem to be present at the start of the optical/$\gamma$-ray activity,
but it becomes clearer later, in agreement with the source behaviour at mm wavelengths.
In particular, both the X-ray and mm flux densities reach the maximum value at the end of the period\footnote{The X-ray and mm brightening continued also after the end of the period considered in this paper \citep[see][and Wehrle et al. 2013, in preparation]{weh12}.}, 
when the $\gamma$ and optical fluxes are instead low. If confirmed, a mm-X correlation with no time delay would imply that both emissions come from the same jet zone and that the X-ray radiation is at least in part the result of an inverse-Compton process on the mm photons, as the hard X-ray spectrum suggests.

A DCF analysis on the optical/$\gamma$ and mm flux densities indicates a good correlation ($\rm DCF_{peak} \sim 0.8$), with a time lag of the mm flux variations relative to the optical/$\gamma$ ones of $\tau=120$--150 d.
The possible scenario therefore is that the radiation we see comes from an inhomogeneous jet, where the mm and X-ray photons are emitted from a region located
downstream from that producing the optical and $\gamma$-ray radiation.

\begin{figure*}
\centering
\includegraphics[width=12cm]{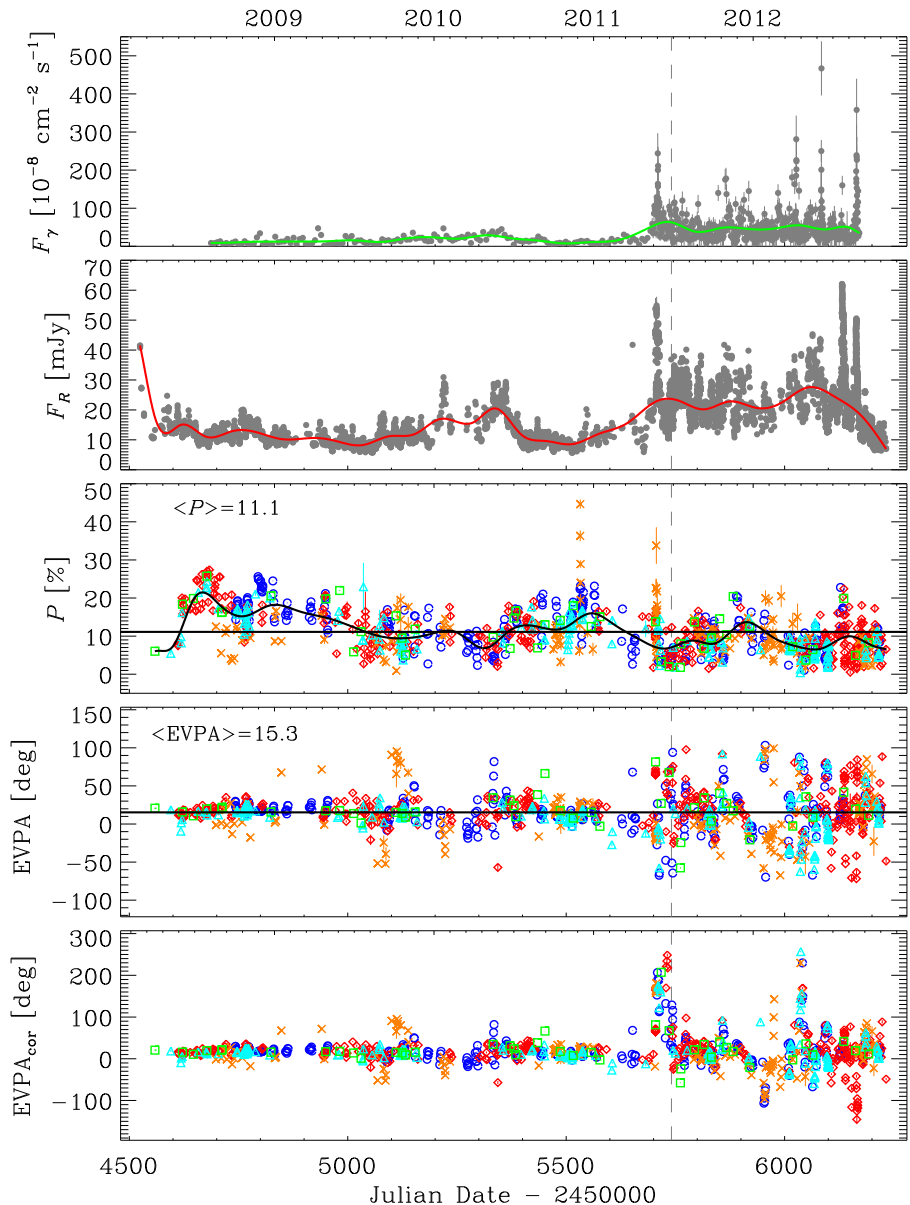}
\caption{Fom top to bottom: a) $\gamma$-ray light curve with cubic spline interpolation through the 60-day binned data (green line);
b) $R$-band light curve with cubic spline interpolation through the 60-day binned data (red line);
c) polarisation percentage with cubic spline interpolation through the 60-day binned data (black line); different symbols and colours refer to different observatories: Calar Alto (green squares), Crimean (red diamonds), Lowell (cyan triangles), Steward (blue circles), and St.\ Petersburg (orange crosses); the horizontal line indicates the average value;
d) electric vector polarisation angle (EVPA); the horizontal line marks its average value;
e) EVPA after correction for the $\pm 180\degr$ ambiguity (see text for details).
In all panels the dashed vertical line indicates the time of the very rapid TeV flare detected by VERITAS.}
\label{pola}
\end{figure*}

\section{Polarisation}

Blazar emission is characterised by variable degree of linear polarisation $P$ and electric vector polarisation angle \citep[EVPA;][]{smi96}. 
In several cases the EVPA was observed to undergo wide rotations during active phases. This is also the case of BL Lacertae, both in the radio \citep{all81} and optical \citep{sil93,mar08} bands.

Photopolarimetric observations of BL Lacertae were performed by \citet{sil93} in 1989, during an outburst, and in 1990. They noticed considerable variability of both $P$ and EVPA during the outburst and suggested two possible interpretations: a jet pointing nearly towards us with helical magnetic field, 
or the interplay of a stable jet component with a linearly rotating component.

An analysis of the long-term (1969--1991) optical polarisation behaviour of BL Lacertae was presented by \citet{hag02}. They found a preferred polarisation direction at EVPA $\approx 20\degr$ and that $P$ in general was higher when the flux was lower and the EVPA was near the preferred value. They interpreted the polarisation variability as due to the superposition of new components with randomly distributed polarisation directions on a persistent, underlying source of polarised radiation with $P=9.2\%$ and EVPA $=24\degr$. The new components lead to a flux increase, but their different EVPAs make the polarisations cancel one another.

When analysing the optical polarimetric behaviour of BL Lacertae around the late 2005 outburst, \citet{mar08} discovered a rotation of the EVPA of 240\degr, in the middle of which the degree of polarisation dropped to a minimum. 
They inferred that the event was caused by the propagation of a shock wave down the jet along a spiral streamline.

We collected 1014 polarisation data in the $R$-band from the Calar Alto, Crimean, Lowell (Perkins), Steward (Bok and Kuiper), and St.\ Petersburg observatories. Details on the data acquisition and reduction procedures can be found in \citet{jor10}, \citet{lar08}, and \citet{smi03}.
Figure \ref{pola} shows $P$ and EVPA compared to both the $R$-band flux densities and 0.1--100 GeV $\gamma$-ray fluxes. Cubic spline interpolations through the 60-day binned light and polarisation curves are drawn to highlight the long-term behaviour. They show that in average $P$ is slowly decreasing during the whole period and that it is higher when the optical and $\gamma$-ray fluxes are low, as found by \citet{hag02}.
The mean value of $P$ is about 11\%, but with variations between 0.4\% and 45\%. In particular, the highest values of $P$ are reached during a very fast spike on $\rm JD=2455532$ that has no counterpart in either optical or $\gamma$-ray flux. Another polarisation peak occurred on $\rm JD=2455706$, a few days after the onset of the optical and $\gamma$-ray outburst of 2011--2012.

The analysis of the EVPA is complicated by the $\pm 180\degr$ ambiguity. To solve for this we proceeded as follows. We first assembled the various datasets asking that all angles were comprised between $-90\degr$ and $+90\degr$. We then calculated the average polarisation angle, $<$EVPA$>$, and iterated the data assemblage by asking that all points are between $<$EVPA$>- \, 90\degr$ and $<$EVPA$>+ \, 90\degr$ until we reach a stable value of $<$EVPA$>$, which is about $15.3\degr$. The resulting angles are plotted in the fourth panel of Fig.\ \ref{pola}.
Our average optical EVPA is very similar to the VLBA radio core EVPA of $13\degr$ estimated by \citet{lis11}. According to the same authors, the mean VLBA jet direction of BL Lacertae is $-171\degr$, in agreement with earlier VLBI results ($-170\degr$) by \citet{gab00}. This means that the optical EVPA is nearly aligned with the radio core EVPA and jet direction.

Spurious jumps of the EVPA due to the $\pm 180\degr$ ambiguity may be corrected by requiring that whenever subsequent points that are separated by less than $\Delta t$ imply angular variations greater than $\Delta$EVPA, they can be shifted by $\pm 180\degr$ in order to minimise the variation. 
The choice of $\Delta t$ and of $\Delta$EVPA is rather arbitrary. From the analysis of \citet{mar08}, we know that we can have variations as large as about 50\degr a day, so we applied the $\pm 180\degr$ correction when $\Delta$EVPA $/ \Delta t > 50\degr/{\rm d}$. The EVPA plot in the bottom panel of Fig.\ \ref{pola} is the result of this procedure. 

\begin{figure}
\centering
\resizebox{\hsize}{!}{\includegraphics{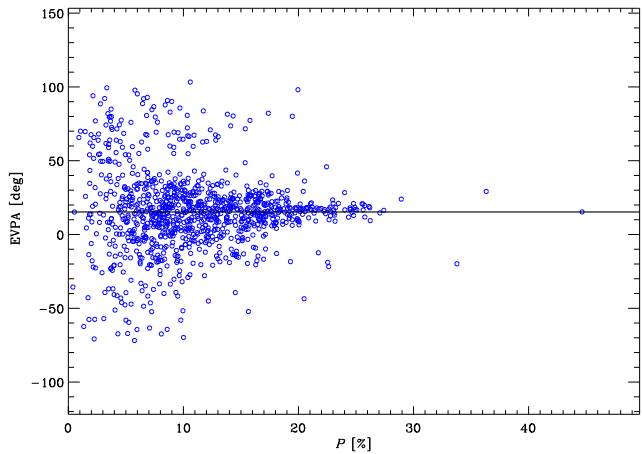}}
\caption{Electric vector polarisation angle (EVPA) as a function of the percentage of polarisation, $P$. The points cluster around a mean EVPA value of 15.3\degr.}
\label{tepi}
\end{figure}

We notice that points cluster around the mean value (see also Fig.\ \ref{tepi}) and that the dispersion around the mean is small before the onset of the 2011--2012 outburst, while the EVPA undergoes much wider changes during the outburst, and this is a general feature, independent of the adopted solution for the $\pm 180\degr$ ambiguity.

During the very fast spike in $P$ on $\rm JD=2455532$ the polarisation angle was close to its mean value, 
but this is not the case for the $\rm JD=2455706$ event, which was preceded by a rotation of the EVPA of about 180\degr\ in $\sim 2$ d. This noticeable EVPA variation occurred at the start of the optical outburst. The other large EVPA rotations (around $\rm JD = 2455730$, $\rm JD=2455975$, and $\rm JD=2456040$) do not correspond to optical or $\gamma$-ray events, and are most likely spurious effects, produced by our arbitrary way of treating the $\pm 180\degr$ ambiguity.

A very rapid TeV flare was detected by the Very Energetic Radiation Imaging Telescope Array System (VERITAS) on 2011 June 28 \citep{arl13}. This was accompanied by changes of the radio and optical polarisation angles and was associated with the emergence of a new superluminal knot in the VLBA radio maps.
From our data we notice that at the time when the rapid TeV $\gamma$-ray flare was detected by VERITAS (2011 June 28, $\rm JD = 2455740.95$),
a change in the EVPA of about $90\degr$ in 1 d was observed, from $40\degr$ on $\rm JD=2455740.88$ to $71\degr$ on $\rm JD=2455741.50$, and to $129\degr$ on $\rm JD=2455741.89$.
The event was also preceded by a small optical flare (36.86 mJy on $\rm JD=2455740.48$), and was followed by a fast jump in polarisation degree (from about 4\% on $\rm JD=2455740.88$ to 12\% on $\rm JD=2455741.50$ and then back to 4\% on $\rm JD=2455741.89$). In contrast, the GeV $\gamma$-ray flux did not show appreciable variations in the same period.

\begin{figure}
\centering
\resizebox{\hsize}{!}{\includegraphics{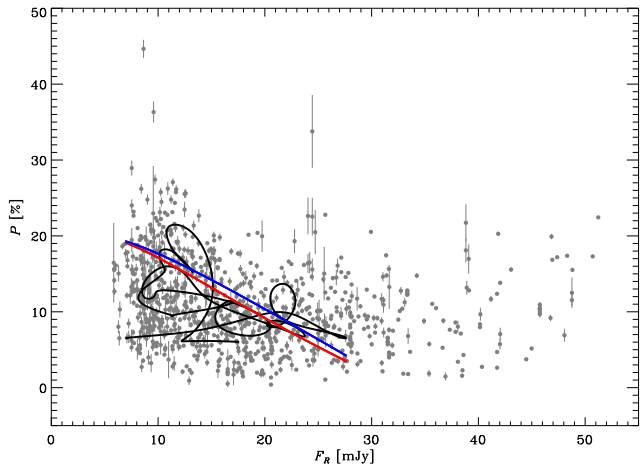}}
\caption{Degree of polarisation as a function of the $R$-band flux density. The tangled line refers to the long-term trend, represented by the cubic spline interpolations through the 60-day binned $F_R$ and $P$ curves shown in Fig.\ \ref{pola}. The blue and red lines represent the results of the helical magnetic field and transverse shock wave models shown in Fig.\ \ref{lyu}. In the latter case a degree of compression of the shock wave $\eta \approx 1.314$ has been chosen.}
\label{hughes}
\end{figure}

In Fig.\ \ref{hughes} we show $P$ as a function of the $R$-band flux density; in general, the plot confirms the trend of a decreasing polarisation with increasing flux noticed above. In particular, this behaviour is marked by the tangled black line, which was obtained by the cubic spline interpolations to the $F_R$ and $P$ data shown in Fig.\ \ref{pola} and thus represents the long-term trend.  However, the data scatter is large, and there are several points with polarisation higher than 15\% and flux density larger than 40 mJy.

\subsection{Near-infrared polarimetry}

In 2011 October, few near-infrared polarimetric observations of BL Lacertae were collecteed using the instrument LIRIS \citep{man04} attached at the 4.2 m William Herschel Telescope (La Palma). LIRIS is a near-infrared public instrument with imaging and spectroscopy capabilities. Polarimetry mode is based on the use of  WeDoWo devices \citep{oli97}. Data were reduced using a dedicated package developed within IRAF ({\tt lirisdr}). The observations of BL Lacertae were part of a more extense program gathering near-infrared polarimetry of a sample of blazars.

\begin{figure}
\centering
\resizebox{\hsize}{!}{\includegraphics{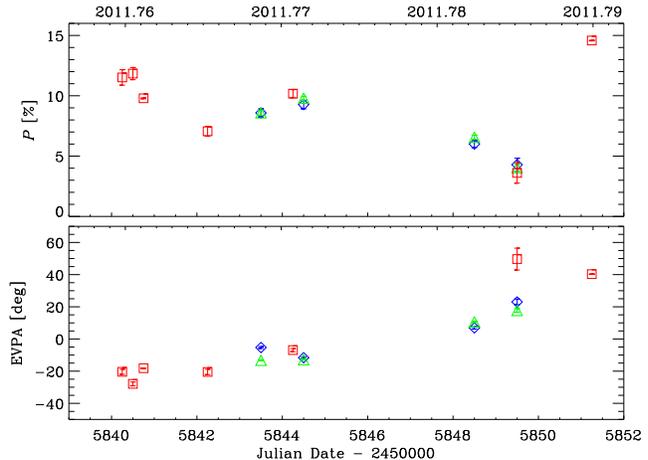}}
\caption{Degree of polarisation $P$ (top) and electric vector polarisation angle (EVPA, bottom) in the period of the near-infrared observations with LIRIS. Blue diamonds and green triangles represent LIRIS data in $J$ and $Ks$ bands, respectively. Red squares show the optical, $R$-band data presented in Fig.\ \ref{pola}.}
\label{liris}
\end{figure}

Figure \ref{liris} shows the polarisation percentage (top) and the EVPA (bottom) as a function of time in the period of the LIRIS observations, comparing the LIRIS $J$ and $Ks$ data to those in the $R$ band.
There is a good agreement between the measurements in bands $J$ and $Ks$, and they also seem to fit the trend of $P$ and EVPA traced by the optical data. The only remarkable difference is the EVPA at $\rm JD=2455849$, where the optical point is about 30\degr\ far from the near-infrared point, but with a large uncertainty.
We conclude that no significant wavelength-dependence of $P$ was detected in the considered period, with the exception of an EVPA measurement.

\section{Geometrical interpretations of the flux and polarisation variability}

Evidence suggests that the relativistic jets in blazars are not straight and steady structures \citep[e.g.][]{kel04,mar11,blo13}. Indeed, instabilities may cause bends in the jet. The jet may rotate because it is tied to the central black hole or accretion disc, or because the central engine is a binary black hole system \citep{vil98a}, and thus it may assume a rotating helical structure \citep{vil99}. As a consequence, we may expect that different emitting regions in the jet have different alignments with the line of sight, which can change in time. Because of the relativistic plasma motion, these changing viewing angles by themselves imply variability (by different amounts at different frequencies), even in the absence of intrinsic flux changes. 

Indeed, the emission from a relativistic plasma is Doppler boosted, so that the observed flux density
$F_\nu(\nu)=\delta^{n+\alpha} F'_{\nu'}(\nu)$, where primed quantities refer to the jet rest frame, $\alpha$ is the intrinsic spectral index, $F'_{\nu'}(\nu') \propto (\nu')^{-\alpha}$, and $n=2$ for a smooth, continuous jet \citep[e.g.][]{urr95}. 
The Doppler factor, $\delta=[\Gamma_{\rm b}(1-\beta \cos \theta)]^{-1}$, depends on both the bulk Lorentz factor of the plasma, $\Gamma_{\rm b}=(1-\beta^2)^{-1/2}$, where $\beta$ is the flow velocity normalised to the speed of light, and the viewing angle $\theta$. Therefore the observed flux can show variability if $\Gamma_{\rm b}$ or $\theta$ change, even if the intrinsic flux remains steady. 
In several blazar studies \citep[e.g.][]{vil02,ost04,rai11,rai12}, we investigated the consequences of assuming that at least the long-term flux variability may be due to geometrical reasons.
We imagine that the emitting jet is a dynamic structure, where different emitting regions can have different orientations with respect to the line of sight, which can also change in time. 

The long-term trend of the optical flux density of BL Lacertae can be represented by the cubic spline interpolation through the 60-day binned $R$-band light curve shown in Fig.\ \ref{pola}.
Adopting $\alpha=1$, we can derive the Doppler factor by $\delta=\delta_{\rm max} (F/F_{\rm max})^{1/3}$, where $\delta_{\rm max}$ is obtained by fixing $\Gamma_{\rm b}=7$ from \citet{jor05} and $\theta_{\rm min}=2\degr$ from \citet{lar10}.
The behaviour of $\delta$ is shown in Fig.\ \ref{lyu}; its value ranges from 8.3 to 13.2.
From the definition of $\delta$ we can then derive $\theta$, also shown in Fig.\ \ref{lyu}. It oscillates between the assumed minimum value of 2\degr\ and 6.8\degr.

\subsection{Helical magnetic fields}

\begin{figure}
\centering
\resizebox{\hsize}{!}{\includegraphics{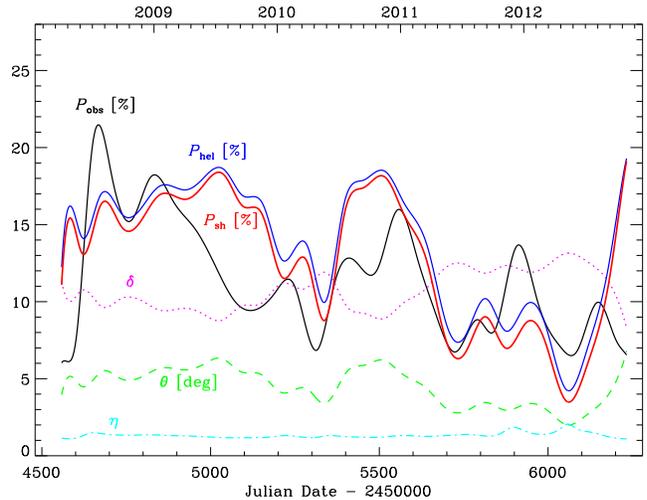}}
\caption{The Doppler factor $\delta$ (dotted pink line) and viewing angle $\theta$ (dashed green line) characterising the optical emission region according to a geometrical interpretation of the long-term optical flux variability.
The black line is the cubic spline interpolation through the 60-day binned observed polarisation curve shown in Fig.\ \ref{pola}. The $P_{\rm hel}$ (blue) and $P_{\rm sh}$ (red) lines represent the long-term polarisation behaviour predicted by the helical magnetic field and transverse shock wave models, respectively.
The dot-dashed cyan line traces the evolution of the degree of compression of the shock wave, $\eta$, which would perfectly reproduce $P_{\rm obs}$.}
\label{lyu}
\end{figure}

We now investigate what would be the implications of this geometric variability scenario on the observed polarisation. 
\citet{lyu05} calculated the polarisation for optically thin synchrotron emission from relativistic jets with helical magnetic fields.
For some of the jet structures they examined\footnote{We refer to the diffuse and reverse-field pinch cases, with number density of relativistic particles scaling according to the square of the intrinsic magnetic field (see Figs.\ 11c and 12c in \citealt{lyu05}).}, the behaviour of the polarisation degree can be approximated as 
\begin{equation}P=P_{\rm max} \sin^2 \theta',
\end{equation}
with $P_{\rm max} \approx 20\%$.
The angle $\theta'$ is the viewing angle in the jet rest frame, which is related to the observed angle $\theta$ through the Lorentz transformation
\begin{equation}\sin \theta'= {{\sin \theta} \over {\Gamma_{\rm b} (1-\beta \cos \theta)}}.
\end{equation}
Figure \ref{lyu} shows the polarisation behaviour predicted by this model, $P_{\rm hel}$, compared with the long-term behaviour of the observed polarisation, $P_{\rm obs}$, represented by the cubic spline interpolation through the 60-day binned polarisation curve shown in Fig.\ \ref{pola}.
Although the agreement is not perfect, it is impressive how the model prediction can reproduce the level of observed polarisation and the main variations without introducing any free parameter. 
In particular, the amplitude of variation is $\Delta P \sim 15\%$ in both cases.
This model also implies a decreasing $P$ for increasing $F_R$ (see Fig.\ \ref{hughes}).

We considered the consequences of varying the value of $\Gamma_{\rm b}$ inside the uncertainty given by \citet{jor05} ($\pm 1.8$) and of small changes in $\theta_{\rm min}$.
Lowering $\Gamma_{\rm b}$ or $\theta_{\rm min}$ would amplify the variations toward low $P$ values, while higher values of $\Gamma_{\rm b}$ or $\theta_{\rm min}$ would produce higher degrees of polarisation with a smaller range of variability.

\subsection{Transverse shock waves}

In \citet{rai12} we analysed the polarisation behaviour of the FSRQ 4C 38.41, adopting the transverse shock wave model by \citet{hug85}, coupled with the geometrical interpretation of the flux variations described above.

Transverse shock waves propagate downstream the jet, affecting the observed polarisation as
\begin{equation}
P \approx P_0~\frac{(1-\eta^{-2})\sin^2\theta'}{2-(1-\eta^{-2})\sin^2\theta'},
\end{equation} 
where $P_0=(\alpha+1)/(\alpha+5/3)$ is the synchrotron polarisation due to a relativistic electron population with particle distribution ${\rm d}N/{\rm d}E \propto E^{-p}$, with $p=2 \, \alpha+1$.
The parameter $\eta$ is the degree of compression of the shock wave.
The angle $\theta'$ is the rest-frame angle between the line of sight and the compression axis, which coincides with the jet axis for transverse shocks, and is subject to the Lorentz transformation mentioned in the previous section.

Figure \ref{lyu} shows that for $\eta \approx 1.314$ (the value that produces the same $P_{\rm max}$ of 20\% as the helical magnetic field model, see Fig.\ \ref{modello}), the polarisation predicted by the shock model, $P_{\rm sh}$, is very similar to $P_{\rm hel}$. 
The corresponding behaviour of $P$ as a function of $F_R$ is shown in Fig.\ \ref{hughes}.
A better agreement between the observed and predicted polarisation can obviously be obtained by changing the parameter $\eta$ in time, i.e.\ assuming that the optical emitting region is crossed by shocks of different strength. In Fig.\ \ref{lyu} we show the time evolution of $\eta$ that would allow the shock model to perfectly reproduce $P_{\rm obs}$.
The range of $\eta$ variation is 1.10--2.05.

Choosing a shorter time binning interval for the long-term trend would produce more oscillations in both the observed and predicted polarisation evolution, without changing the general scenario.

A more sophisticated application of the \citet{hug85} model to the optical photometric and polarimetric observations of another BL Lac object, S5 0716+71, during the 2011 outburst was performed by \citet{lar13}.
They successfully interpreted the general multifrequency behaviour of the outburst assuming a shock wave propagating along a helical path in the blazar jet.

\subsection{Comparison between BL Lacertae and 4C 38.41}

\citet{rai12} showed that the degree of polarisation of the FSRQ 4C 38.41 increases with the optical flux, also after the unpolarised component likely due to thermal emission from the accretion disc is subtracted.
This was explained by transverse shock waves travelling inside the jet, adopting a high bulk Lorentz factor $\Gamma_{\rm b}=31.1$.
In contrast, the shock model applied to BL Lacertae can explain the observed anticorrelation between the long-term polarisation and flux (Fig.\ \ref{hughes}).
This is due to the much lower Lorentz factor $\Gamma_{\rm b}=7$ used for this object, which implies a less dramatic aberration of the viewing angle.

\begin{figure}
\centering
\resizebox{\hsize}{!}{\includegraphics{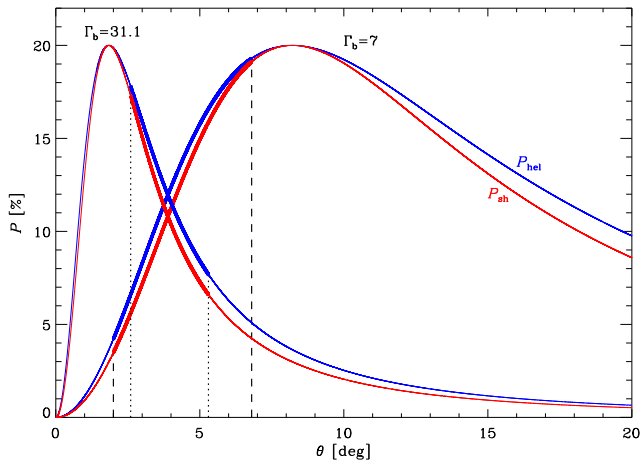}}
\caption{The helical magnetic field (blue lines) and transverse shock wave (red lines) polarisation models for two values of the bulk Lorentz factor. The degree of compression of the shock wave $\eta$ was fixed to $\approx 1.314$ to have the same normalisation as in the helical magnetic field model. The case $\Gamma_{\rm b}=31.1$ refers to the quasar-type blazar 4C 38.41 \citep{rai12}, while $\Gamma_{\rm b}=7$ represents BL Lacertae. The thick portions of the lines mark the range of viewing angle $\theta$ spanned by the optical emitting region according to a geometrical interpretation of the long-term flux variability. The range is in the descending part of the model curves for 4C 38.41, implying smaller $P$ for larger $\theta$, i.e. lower flux. In contrast, for BL Lacertae the range of $\theta$ values is on the ascending portion of the model curves, leading to an anticorrelation between the polarisation and flux.}
\label{modello}
\end{figure}

In Fig.\ \ref{modello} we plotted both $P_{\rm hel}$ and $P_{\rm sh}$\footnote{We choose $\eta \approx 1.314$ for the shock model to have the same normalisation of $P_{\rm hel}$, i.e. $P_{\rm max}=20$.} as a function of the viewing angle $\theta$, for $\Gamma_{\rm b}=31.1$ (4C 38.41 case) and $\Gamma_{\rm b}=7$ (BL Lacertae case). We notice how similar the two models are, $P_{\rm hel}$ being slightly higher at a given $\theta$.
Starting from $\theta=0$ (perfect alignment with the line of sight), the polarisation first grows, reaches a maximum at $\theta \sim 1/\Gamma_{\rm b} \, \rm rad$, and then more slowly decreases. In the case of 4C 38.41, the peak of polarisation occurs at $\theta \approx 1.84\degr$, while for BL Lacertae at $\theta \approx 8.21\degr$. The geometrical interpretation of the long-term optical flux variability of 4C 38.41 led \citet{rai12} to infer that the viewing angle of the corresponding emission region varied between 2.6\degr\ and 5.3\degr. From Fig.\ \ref{modello} we can see that this $\theta$ range is on the descending part of the model curves, so that increasing $\theta$, i.e.\ reducing $\delta$ and hence the flux, the polarisation diminishes.
The long-term trend for BL Lacertae implies a change of the viewing angle from 2.0\degr\ to 6.8\degr. This range is on the rising part of the corresponding model curves, so that an increase in the viewing angle leads to a growth of the polarisation.

\section{Conclusions}

We have analysed the behaviour of BL Lacertae during 2008--2012 at millimetre, optical, UV, X-ray, and $\gamma$-ray frequencies. A general correlation is found between the optical and $\gamma$ flux variations, which are consistent with being simultaneous, suggesting that the observed optical and $\gamma$-ray photons are produced in the same jet region. The $\gamma$-ray flux variation roughly goes as the square of the optical one, suggesting that $\gamma$-ray photons are produced by inverse-Compton scattering of the low-energy synchrotron photons off their parent relativistic electrons (SSC mechanism).
The behaviour of the X-ray flux seems to trace that at mm wavelengths, whose variations follow those at optical/$\gamma$-ray energies by about $\tau \sim 120$--150 d. This implies that the mm and X-ray observed radiation comes from a jet zone that is located downstream the optical/$\gamma$ emitting region. 
The distance between the two emitting zones can be estimated as $D \sim \beta \, c \, \Gamma \, \delta \, \tau /(1+z)$.
Assuming $\Gamma \sim 7$ and $\delta \sim 10$ as adopted/derived in this paper, $D$ ranges from 6.5 to 8.2 pc. This means that the mm/X-ray emitting region is located far away from the AGN central engine, outside the broad line region, which extends on sub-parsec scales. Therefore, as in the case of the $\gamma$-ray radiation, also the X-ray photons are more likely produced by an SSC process. The alternative possibility would be that the seed photons for the inverse-Compton scattering come from a dusty torus. Following \citet{nen08}, the torus external radius can be estimated as  $R_{\rm ext} < 12 \, \sqrt{L_{\rm disc}/(10^{45} \, \rm erg \, s^{-1})} \, \rm pc$ which, for a disc luminosity of $L_{\rm disc} \ga 6 \times 10^{44} \, \rm erg \, s^{-1}$ as derived by \citet{rai09}, gives $R_{\rm ext} \la 10 \, \rm pc$.
As a consequence, even if the distance of the optical/$\gamma$ zone from the black hole were negligible, the mm/X-ray emitting region would be located at the outer bound of the torus. In any case, this picture is questioned by the lack of observable torus emission in BL Lac objects \citep{plo12}.

A more detailed study of the cross-correlation between different bands is severely limited by even small gaps in the data sampling, because of the extremely rapid variability of the source flux.
Optical flares seem to last longer than the corresponding $\gamma$ events, maybe because they are a convolution of many more events. Indeed, the fact that the optical light curve appears as more structured than the $\gamma$-ray curve can only be partially explained by the different sampling. One possible explanation is that the optical emitting region itself presents substructures \citep{nar12}, and that not all of them produce $\gamma$-ray photons.

We have suggested a geometrical interpretation of the long-term flux variability, where different emission regions in the jet have different orientations with respect to the line of sight, which can change over time. 
These orientation changes lead to observed flux variations even when the intrinsic flux does not vary. In particular, the viewing angle $\theta$ of the zone producing the optical photons should vary between 2\degr\ and 6.8\degr\ to explain the long-term trend of the optical flux in the considered period. We have analysed the consequences of this variable orientation on the evolution of the mean optical polarisation. We have found that the helical magnetic field model by \citet{lyu05}, where $P$ is a function of $\theta$ only, naturally generates changes in $P$ in the same range as that observed and reproduces the main observed variations. The fact that the model prediction is not able to match the long-term polarisation curve in detail suggests that this interpretation is too simple and that there is something else that we must take into account.
A possible solution comes from the transverse shock wave model by \citet{hug85}.
This model gives very similar results to those of the helical magnetic field model for a given choice of the degree of compression of the shock wave. 
If we assume that shock waves of different strength can travel down the jet, then the observed long-term trend of $P$ can be fully explained. 

When coupled with the geometrical interpretation of the flux variability, these models offer a simple explanation for the observed correlation/anticorrelation between the long-term polarisation and flux in different sources, which appears to depend on the bulk Lorentz factor for any given range of viewing angles.

\section*{Acknowledgments}
We thank the referee, Philip Hughes, for useful comments.
V.T.Doroshenko acknowledges support by the grant of the Russian Foundation for Basic Research 12-02-01237-a.
Data from the Steward Observatory spectropolarimetric monitoring project were used. This program is supported by Fermi Guest Investigator grants NNX08AW56G, NNX09AU10G, and NNX12AO93G. 
This article is partially based on observations made with the IAC80 operated on the island of Tenerife by the Instituto de Astrofísica de Canarias in the Spanish Observatorio del Teide. Many thanks are due to GAS (IAC Support Astronomer Group) and IAC telescope operators for helping with the observations at IAC-80 telescope.
The Abastumani team acknowledges financial support of the project FR/639/6-320/12 by the Shota Rustaveli National Science Foundation under contract 31/76.
This research was partially supported by Scientific Research Fund of the Bulgarian Ministry of Education and Sciences under grant DO 02-137 (BIn-13/09). The Skinakas Observatory is a collaborative project of the
University of Crete, the Foundation for Research and Technology -- Hellas, and the Max-Planck-Institut f\"ur Extraterrestrische Physik.
St.Petersburg University team acknowledges support from Russian RFBR foundation via grants
09-02-00092 and 12-02-31193-2.
The research at Boston University (BU) was funded in part by NASA Fermi Guest Investigator grants NNX08AV65G, NNX10AO59G, NNX10AU15G, NNX11AO37G, and NNX12AO90G.
The PRISM camera at Lowell Observatory was developed by K.\ Janes et al. at BU and Lowell Observatory, with funding from the NSF, BU, and Lowell Observatory. The Liverpool Telescope is operated on the island of La Palma by Liverpool John Moores University in the Spanish Observatorio del Roque de los Muchachos of the Instituto de Astrofisica de Canarias, with funding from the UK Science and Technology Facilities Council.
This paper is partly based on observations carried out at the German-Spanish Calar Alto Observatory, which is jointly operated by the MPIA and the IAA-CSIC.
This paper is partly based on observations carried out at the IRAM-30 m Telescope, which is supported by INSU/CNRS (France), MPG (Germany), and IGN (Spain).
Acquisition and reduction of the MAPCAT and IRAM 30m data is supported in part by MINECO (Spain) grant and AYA2010-14844, and by CEIC (Andaluc\'{i}a) grant P09-FQM-4784.
The Submillimeter Array is a joint project between the Smithsonian Astrophysical Observatory and the Academia Sinica Institute of Astronomy and Astrophysics and is funded by the Smithsonian Institution and the Academia Sinica.
We acknowledge the use of public data from the $Swift$ data archive.
This research has made use of 
\begin{itemize}
\item the XRT Data Analysis Software (XRTDAS) developed under the responsibility of the ASI Science Data Center (ASDC), Italy;
\item the NASA's Astrophysics Data System Bibliographic Services (ADS);
\item the NASA/IPAC Extragalactic Database (NED) which is operated by the Jet Propulsion Laboratory, California Institute of Technology, under contract with the National Aeronautics and Space Administration. \item data obtained through the High Energy Astrophysics Science Archive Research Center Online Service, provided by the NASA/Goddard Space Flight Center.
\end{itemize}

\bsp

\label{lastpage}

\end{document}